\documentclass[12pt]{article}\usepackage{xoracles}
\input{MyQcircuit.sty}
\begin{document}

\title{
Oracular Approximation of\\
Quantum Multiplexors and \\
Diagonal Unitary Matrices}

\author{
Robert R. Tucci\\
P.O. Box 226\\
Bedford,  MA   01730\\
tucci@ar-tiste.com}
\date{ \today}

\maketitle

\vskip .01 cm
\section*{Abstract}
We give a new quantum circuit approximation
of quantum multiplexors based
on the idea of complexity theory oracles.
As an added bonus, our
multiplexor approximation
immediately gives a
quantum circuit  approximation
of diagonal unitary matrices.

\newpage
\section*{}
For an explanation of the notation
used in this paper,
see Ref.\cite{TucMetHas} Section 2.

Quantum multiplexors have proved themselves
to be  very
useful as building
blocks for quantum
computing circuits. For a review of
quantum multiplexors,
see Ref.\cite{TucMetHas} Section 3.

As shown in Ref.\cite{TucCompiler},
an $R_y(2)$-multiplexor with
$N_B$ controls can be compiled exactly
using $2^{N_B}$ CNOTs.\footnote{
If we
express a quantum circuit as
a sequence of single-qubit rotations
and CNOTs, then the number
of CNOTs can be used as a measure of the
time complexity of the circuit.
Being two-body interactions,
CNOTs take much longer to perform
physically than single-qubit rotations, so
we only count the former.}
It is believed that this number of CNOTs is
a lower bound.
It is therefore of interest
to find
multiplexor approximations
 with a lower CNOT count.
Various multiplexor
approximations have been
considered before\cite{TucMuxorApprox}.
The goal of this paper is to give
a new multiplexor approximation
based on the idea of
complexity theory oracles.
As we shall see, any
multiplexor approximation
immediately gives an approximation
of diagonal unitary matrices.
Diagonal unitary matrices
of dimension $2^\nb$
can also be compiled exactly
using about $2^\nb$ CNOTs\cite{TucCompiler},
and this number is believed to be a lower bound.

Consider an arbitrary $R_y(2)$-multiplexor
whose target qubit is labelled $\tau$
and whose $N_\beta$ control qubits are labelled
$\vec{\beta}=
(\beta_{N_\beta-1},\ldots, \beta_1, \beta_0)$.

\begin{subequations}\label{eq-def-m}
\begin{eqnarray}
M &=& \exp\left(i\sum_{\vecb\in Bool^{N_\beta}}
\theta_\vecb \sigy(\tau) P_\vecb(\vec{\beta})\right)
\\
&=&
\sum_{\vecb\in Bool^{N_\beta}}
e^{i
\theta_\vecb \sigy(\tau) }
P_\vecb(\vec{\beta})
\\
&=&\prod_{\vecb\in Bool^{N_\beta}}
\exp\left(i
\theta_\vecb \sigy(\tau) P_\vecb(\vec{\beta})\right)
\;,
\end{eqnarray}
\end{subequations}
for some $\theta_\vecb\in \RR$.
In the above,
we define $\vecb=(b_{N_\beta-1},\ldots, b_1, b_0)$,
$\vec{\beta}=(\beta_{N_\beta-1},\ldots, \beta_1, \beta_0)$,
and
$P_\vecb(\vec{\beta})=
\prod_{j=0}^{N_\beta-1} P_{b_j}(\beta_j)$.
Also,
$P_0 = \ket{0}\bra{0}=\nbar = 1-n$
and $P_1 = \ket{1}\bra{1}= n$,
where $n$ is the so called ``number operator".
In Eqs.(\ref{eq-def-m}), we've expressed $M$ in 3 equivalent forms,
the exponential, sum and product forms.
The equivalence of these forms is readily established
by applying $\ket{\vecb}_{\vec{\beta}}$ to
 the right hand side
of each form.
Note that
we can ``pull" the $\vecb$ sum out of the exponential,
but only if we also pull out the projector $P_\vecb$.

\begin{figure}[h]
    \begin{center}

\beq
\begin{array}{c}
\Qcircuit @C=1em @R=1em @!R{
\lstick{\beta_0}
&\muxorgate
&\qw
&
&
&
&\multigate{2}{\;}
&\multigate{2}{\;}
&\multigate{2}{\;}
&\qw
&\qw
&\qw
&\multigate{2}{\;}
&\multigate{2}{\;}
&\multigate{2}{\;}
&\qw
\\
\lstick{\beta_1}
&\muxorgate
&\qw
&
&
&
&\ghost{\;}
&\ghost{\;}
&\ghost{\;}
&\qw
&\qw
&\qw
&\ghost{\;}
&\ghost{\;}
&\ghost{\;}
&\qw
\\
\lstick{\beta_2}
&\muxorgate
&\qw
&
&
&
&\ghost{\;}
&\ghost{\;}
&\ghost{\;}
&\qw
&\qw
&\qw
&\ghost{\;}
&\ghost{\;}
&\ghost{\;}
&\qw
\\
\lstick{\alpha_1}
&\qw
&\qw
&\ket{0}
&
&\lstick{\approx}
&\timesgate\qwx[-1]
&\qw
&\qw
&\dotgate
&\qw
&\qw
&\qw
&\qw
&\timesgate\qwx[-1]
&\qw
&\rstick{\ket{0}}
\\
\lstick{\alpha_2}
&\qw
&\qw
&\ket{0}
&
&
&\qw
&\timesgate\qwx[-2]
&\qw
&\qw
&\dotgate
&\qw
&\qw
&\timesgate\qwx[-2]
&\qw
&\qw
&\rstick{\ket{0}}
\\
\lstick{\alpha_3}
&\qw
&\qw
&\ket{0}
&
&
&\qw
&\qw
&\timesgate\qwx[-3]
&\qw
&\qw
&\dotgate
&\timesgate\qwx[-3]
&\qw
&\qw
&\qw
&\rstick{\ket{0}}
\\
\lstick{\tau}
&\emptygate\qwx[-6]
&\qw
&
&
&
&\qw
&\qw
&\qw
&\emptygate\qwx[-3]
&\emptygate\qwx[-2]
&\emptygate\qwx[-1]
&\qw
&\qw
&\qw
&\qw
}
\end{array}
\nonumber
\eeq
    \caption{
    Oracular approximation of an
    $R_y(2)$-multiplexor
    with 3 controls, with
    the angles stored to a precision of 3
fractional bits.
    }
    \label{fig-m-approx}
    \end{center}
\end{figure}
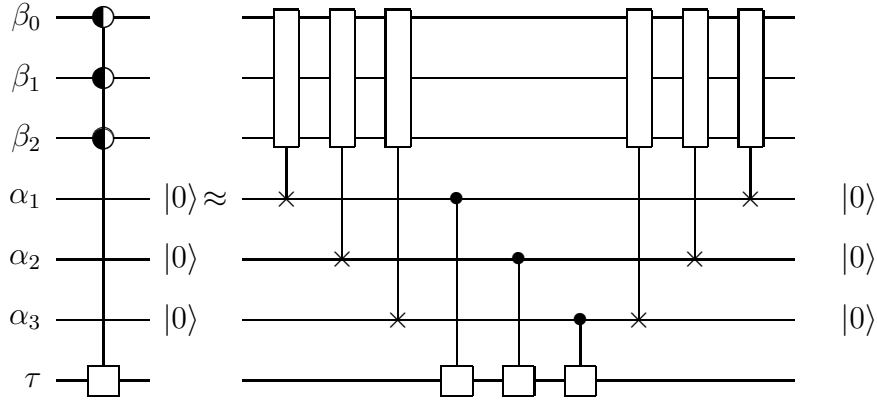

Now we
add a set of $N_\alpha$ ancilla qubits
labelled $\vec{\alpha} =(\alpha_1, \alpha_2,
\ldots, \alpha_{N_\alpha})$.
For each $\vecb$,
the angle $\theta_\vecb$
can be expressed approximately,
to a precision of $N_\alpha$
fractional bits,
and this information
can be stored in the
qubits $\vec{\alpha}$.
Let $\ket{0}_{\vec{\alpha}}=
\prod_{k=1}^{N_\alpha}
\ket{0}_{\alpha_k}$. If

\beq
\theta_\vecb
= 2\pi \sum_{k=1}^{N_\alpha}
\frac{a_{\vecb, k}}{2^k}
\;,
\eeq
where $a_{\vecb, k}\in Bool$,
then

\beq
\theta_\vecb\ket{0}_{\vec{\alpha}}
=
\left[
\prod_{k=1}^{N_\alpha}
\sigx(\alpha_k)^{a_{\vecb, k}}
\right]
2\pi \sum_{k=1}^{N_\alpha}
\frac{n(\alpha_k)}{2^k}
\left[
\prod_{k=1}^{N_\alpha}
\sigx(\alpha_k)^{a_{\vecb, k}}
\right]
\ket{0}_{\vec{\alpha}}
\;.
\label{eq-theta-in}
\eeq
We will use the following evocative notation
for the finite series:
\beq
\sum_{k=1}^{N_\alpha}
\frac{n(\alpha_k)}{2^k}
=
0.n(\alpha_1)n(\alpha_2)
\ldots n(\alpha_{N_\alpha})
\;.
\label{eq-cute-notation}
\eeq
Substituting Eqs.(
\ref{eq-cute-notation})  into Eq.(\ref{eq-theta-in}),
and then using the resulting expression
for $\theta_\vecb\ket{0}_{\vec{\alpha}}$ in the definition Eq.(\ref{eq-def-m})
for $M$, gives:

\beqa
 {\scriptstyle M(\vec{\beta}, \tau)
 \ket{0}_{\vec{\alpha}}}
 &=&
{\scriptstyle\sum_{\vecb}P_\vecb(\vec{\beta})
\left[
\prod_{k=1}^{N_\alpha}
\sigx(\alpha_k)^{a_{\vecb, k}}
\right]
e^{i 2\pi\;
0.n(\alpha_1)n(\alpha_2)
\ldots n(\alpha_{N_\alpha})
\sigy(\tau)}
\left[
\prod_{k=1}^{N_\alpha}
\sigx(\alpha_k)^{a_{\vecb, k}}
\right]
\ket{0}_{\vec{\alpha}}}\\
&=&
\Omega(\vec{\alpha})
e^{i 2\pi\;
0.n(\alpha_1)n(\alpha_2)
\ldots n(\alpha_{N_\alpha})
\sigy(\tau)}
\Omega(\vec{\alpha})
\ket{0}_{\vec{\alpha}}
\;,
\eeqa
where

\beq
\Omega(\vec{\alpha})= \prod_{k=1}^{N_\alpha}
\Omega(\alpha_k)
\;,
\eeq
where

\beq
\Omega(\alpha_k)=
\sum_\vecb P_\vecb(\vec{\beta})\sigx(\alpha_k)^{ a_{\vecb, k}}
=
\sigx(\alpha_k)^{\sum_\vecb a_{\vecb, k}P_\vecb(\vec{\beta})}
\;.
\eeq
$\Omega(\vec{\alpha})$ is a product of $N_\alpha$
``standard quantum oracles"
$\Omega(\alpha_k)$.\footnote{For
an introduction to quantum
oracles from a quantum computer programmer's perspective,
see Ref.\cite{qbnets-blog}.}
For example, if $N_\alpha=2$ and $N_\beta=2$ with

\beq
\begin{array}{l}
\theta_{00} = 2\pi 0.01\\
\theta_{01} = 2\pi 0.11\\
\theta_{10} = 2\pi 0.10\\
\theta_{11} = 2\pi 0.00
\end{array}
\;,\;\;
[a_{\vecb,k}]
=
\begin{tabular}{|ll||ll}
\hline
&&$k\rarrow$\\
&&\tiny{1}&\tiny{2}\\
\hline\hline
&\tiny{00}&0&1\\
$\vecb$&\tiny{01}&1&1\\
$\downarrow$&\tiny{10}&1&0\\
&\tiny{11}&0&0
\end{tabular}
\;,
\eeq
then

\beq
\Omega(\vec{\alpha}) =
\begin{array}{c}
\Qcircuit @C=1em @R=1em @!R{
&\timesgate
&\timesgate
&\qw
&\qw
&\rstick{\alpha_1}
\\
&\qw\qwx
&\qw\qwx
&\timesgate
&\timesgate
&\rstick{\alpha_2}
\\
&\dotgate\qwx
&\ogate\qwx
&\ogate\qwx
&\dotgate\qwx
&\rstick{\beta_0}
\\
&\ogate\qwx
&\dotgate\qwx
&\ogate\qwx
&\ogate\qwx
&\rstick{\beta_1}
}
\end{array}
\;\;\;\;\;.
\eeq
Fig.\ref{fig-m-approx}
is an example,  expressed in
circuit form, of the
multiplexor approximation that
we hath wrought.

Next, consider
a diagonal unitary matrix $D$ acting on
qubits $\vec{\beta}$:

\beq
D =
\exp\left(i\sum_{\vecb}
\theta_\vecb P_\vecb(\vec{\beta})\right)
\;,
\eeq
for some $\theta_\vecb\in \RR$.
By adding an ancilla target qubit $\tau$,
we can express $D$
in terms of an $R_z(2)$-multiplexor:

\beqa
D(\vec{\beta})\ket{0}_\tau
&=&
\exp\left(i\sum_{\vecb}
\theta_\vecb
\sigz(\tau)P_\vecb(\vec{\beta})\right)
\ket{0}_\tau\\
&=&
e^{-i \frac{\pi}{4}\sigx(\tau)}
\exp\left(i\sum_{\vecb}
\theta_\vecb
\sigy(\tau)P_\vecb(\vec{\beta})\right)
e^{i \frac{\pi}{4}\sigx(\tau)}
\ket{0}_\tau
\;
\eeqa
An oracular approximation of $D$
follows immediately from this
and the oracular multiplexor approximation.

In general, if a
classical algorithm
of polynomial complexity is known
for calculating
$\theta_\vecb$
given $\vecb$,
then one can
construct from this
classical algorithm
standard quantum
oracles $\{\Omega(\alpha_k)\}_{\forall k}$,
of polynomial complexity. {\it  However,
if no such classical
algorithm of polynomial complexity is known,
none may exist. If none exists,
the
standard quantum
oracles $\Omega(\alpha_k)$
have exponential complexity.}

An upper bound on the error
of our oracular multiplexor
approximation is
easily obtained. Let

\beq
\theta = 2\pi\sum_{k=1}^{\infty} \frac{a_k}{2^k}
\;\;,\;
\hat{\theta} = 2\pi\sum_{k=1}^{N_\alpha} \frac{a_k}{2^k}
\;.
\eeq
Then

\begin{subequations}
\begin{eqnarray}
|e^{i\theta}- e^{i\hat{\theta}}|&=&
|e^{i\theta}|\left|1-\exp(-i2\pi
\sum_{k=N_\alpha+1}^{\infty}
\frac{a_k}{2^k}
)\right|
\label{eq-bef-sin-ineq}
\\
&\leq&
2\pi \left|
\sum_{k=N_\alpha+1}^{\infty}
\frac{a_k}{2^k}
\right|
\label{eq-aft-sin-ineq}
\\
&\leq&
2\pi \left|
\sum_{k=N_\alpha+1}^{\infty}
\frac{1}{2^k}
\right|
  =
\frac{2\pi}{2^{N_\alpha}} \left|
\sum_{k=1}^{\infty}
\frac{1}{2^k}
\right|
=
\frac{2\pi}{2^{N_\alpha}}
\;.
\end{eqnarray}
\end{subequations}
To go from Eq.(\ref{eq-bef-sin-ineq})
to Eq.(\ref{eq-aft-sin-ineq}), we used the
inequality
$|1-e^{ix}|
=2|\sin\frac{x}{2}|\leq |x|$ for $x\in \RR$.
Almost the same string
of inequalities can be used
to upper bound the error in
our oracular multiplexor approximation.
Let $M$ be the multiplexor
of Eq.(\ref{eq-def-m}) with
 the exact $\theta_\vecb$,
 and $\hat{M}$
 the multiplexor with the
 approximate  $\theta_\vecb$ (to a precision
 of $N_\alpha$ fractional bits).
Using the matrix 2-norm\footnote{The
2-norm of a matrix
 is defined as its largest singular value (singular
 values defined $\geq$ 0). For a review
 of matrix norms, see Ref.\cite{Golub}},
 we get

\beqa
\| M- \hat{M}\|&=&
\| M \|
\left\| 1-\exp[-i2\pi
\sum_\vecb
\sum_{k=N_\alpha+1}^{\infty}
\frac{a_{\vecb,k}}{2^k}
\sigy(\tau)P_\vecb(\vec{\beta})
]\right\|\\
&\leq&
2\pi \| \sigy(\tau) \| \left\|
\sum_\vecb
\sum_{k=N_\alpha+1}^{\infty}
\frac{a_{\vecb,k}}{2^k}P_\vecb(\vec{\beta})
\right\|
\leq
2\pi \max_\vecb \left\|
\sum_{k=N_\alpha+1}^{\infty}
\frac{a_{\vecb,k}}{2^k}
\right\|
\\
&\leq&
2\pi \left|
\sum_{k=N_\alpha+1}^{\infty}
\frac{1}{2^k}
\right|
  =
\frac{2\pi}{2^{N_\alpha}} \left|
\sum_{k=1}^{\infty}
\frac{1}{2^k}
\right|
=
\frac{2\pi}{2^{N_\alpha}}
\;.
\eeqa
Above, we used the fact that
$\| M\|=
\| \sigy(\tau) \|=1$
as is the case for any unitary matrix.

\section*{Acknowledgements}
I got the idea for this paper from
an exchange of emails between me,
Rolando Somma, Emmanuel Knill, Howard Barnum,
Pawel Wocjan and Richard Cleve.(Everyone
received the same emails).
I offered to one of these individuals
to include him as a co-author
of this paper, but he
refused the offer (didn't even
want to be acknowledged).
None of the above mentioned
people that corresponded with me
has vetted or proof-read this paper.
Any errors are my own.
There is
some indirect precedent to this work.
According to Prof. Wocjan, ``It is known how to efficiently simulate sparse Hamiltonians (for
example, see the papers by Dorit Aharonov, et.al., Sanders, Cleve,
et.al., which Rolando mentioned in his reply to you).  These methods do
not immediately allow to efficiently implement a sparse unitary matrix.
Stephen Jordan and I found a simple method to simulate any sparse matrix
by using the techniques for simulating sparse Hamiltonians.  While this
method is not very difficult, we still want to write a short note
because the ability to simulate sparse unitaries can be used to derive
quantum algorithms for evaluating link invariants ..."
I have not seen the
so far unpublished work of Jordan/Wocjan.
I don't believe
that I use ``the techniques for simulating sparse Hamiltonians"
in this paper.
Furthermore, it appears
that Jordan/Wocjan consider sparse unitary
matrices
in general, whereas I consider
the very special sparse case of quantum
multiplexors. Sometimes, by
considering a special case, one can say
much more and be more specific.
For these reasons, I believe that this paper might
be significantly different to the one
by Jordan/Wocjan, and
that it might
be useful to someone besides me.

\end{document}